\documentclass[twocolumn]{IEEEtran}
\usepackage{epsfig,amsmath,amstext,amsfonts,amssymb,cite,cases,multirow}
\usepackage{algorithm} 
\usepackage{algorithmic} 
\usepackage{mathrsfs}
\usepackage{color}

\usepackage{amsmath}
\usepackage{graphicx}
\usepackage{subfigure}
\usepackage{bm}
\usepackage{subeqnarray}
\usepackage{cases}
\usepackage{diagbox}
\usepackage{makecell}

\begin{document}
	
\title{Numerology Selection for OFDM Systems Based on Deep Neural Networks}

\author{
	Xiaoran~Liu, ~
	Jiao~Zhang,
	and~Jibo~Wei,~\IEEEmembership{Member,~IEEE}
	\thanks{This work was supported in part by the National Natural Science Foundation of China under Grant 61931020 and 61601480.}
	\thanks{X. Liu, J. Zhang and J. Wei are with the College of Electronic Science and Technology, National University of Defense Technology, Changsha 410073, China (e-mail: \{liuxiaoran10, zhangjiao16, wjbhw\}@nudt.edu.cn).	}
}

\maketitle

\begin{abstract}
	In order to support diverse scenarios and deployments, 
	the numerology of orthogonal frequency division multiplexing (OFDM) is defined for the parametrization of subcarrier spacing and cyclic prefix (CP).	
	The time-frequency dispersion of mobile radio channels and 
	the channel noise result in different performance deterioration in different numerologies.	
	In this letter, we propose a deep neutral network (DNN) approach for numerology selection of OFDM systems.	
	Considering the inter-symbol interference (ISI), inter-carrier interference (ICI) and noise level, the SNR loss is established as the objective to be minimized.
	We extract the power delay profile, mobile velocity and noise power as the input features to the DNN.
	The proposed DNN learns from the channel characteristics to obtain the optimal numerology selection.
	Simulation results show that the proposed DNN achieves better performance than the existing methods.
	The decision boundaries of different numerologies are also illustrated to show the application range according to the channel characteristics.
	
\end{abstract}

\begin{IEEEkeywords}
	Numerology selection, orthogonal frequency division multiplexing (OFDM), deep neural networks, channel characteristics.
\end{IEEEkeywords}

\section{Introduction}

%
%
%
Future wireless communications systems are expected to support a wide range of usage scenarios.
Numerologies defined for orthogonal frequency division multiplexing (OFDM) in 5G New Radio (NR) achieve a great improvement on flexibility \cite{2018ICSM-OFDM}. 
The OFDM numerologies with different waveform parameters such as subcarrier spacing and cyclic prefix (CP) length are designed to combat the time and frequency dispersion characteristics of the radio channel that lead to inter-symbol interference (ISI) and inter-carrier interference (ICI).
By adopting different numerologies, the communication systems have the capability to adapt to different channel characteristics.
To be specific, the ISI caused by multipath delay spread can be reduced by increasing the CP length and using large symbol
duration which is realized by applying small subcarrier spacing. 
However, the prolonged CP leads to low power efficiency, and the lengthening of the symbol duration makes the communication systems more sensitive to the Doppler spreads which damage the orthogonality of the subcarriers and bring about more ICI.
Therefore, as different numerologies suffer from distinct levels of interference under the specific channel characteristic,
the numerology selection problem should be carefully considered to guarantee the acceptable signal-to-interference-plus-noise ratio (SINR).

In order to minimize the ISI and ICI, the authors in \cite{2013ITWC-Optimal} propose an optimal subcarrier spacing selection method. By the analysis of the expectation of interference power, an optimization problem is formulated to minimize the total interference.
However, due to the approximation of optimization objective, this method cannot always provide the optimum selection.
Moreover, the reduction of power efficiency caused by the CP overhead is not considered.
The CP ratio is invariable in \cite{2013ITWC-Optimal}, which cannot provide further improvement on the waveform flexibility.
In \cite{2018IA-Flexibility}, the numerology set is significantly extended to increase overall flexibility and a heuristic method is then developed to find the balance between the numerology number and different user requirements. 
The authors further describe a waveform parameter assignment framework using machine learning and provide a prospect of configurable waveform in \cite{2020IOJoVT-Waveform}.
The previous works, however, do not quantitatively consider the effect of the channel characteristics, so that the selected numerology cannot be guaranteed to be the optimum in terms of the ISI and ICI.

Recently, deep learning has achieved great success in image classification task and decision-making, etc.
Based on a cascade of multiple layers of nonlinear processing units, it can be optimized for minimizing classification error.
Notably, due to its excellent nonlinear characteristics, deep learning has been confirmed to be powerful in wireless communications, such as signal identification \cite{2018ICL-Specific} and adaptive modulation coding (AMC)\cite{2019IWCL-Deep}.


Motivated by these existing works, in this paper, we propose a deep neutral
network (DNN)-based approach to select the optimal numerology from the candidate numerology set, which can achieve the minimum SNR loss.
Since the ISI and ICI are caused by the delay and Doppler spreads of the wireless channel and the SNR is dependent on the noise level, we use the power delay profile (PDP), mobile velocity and the noise power as the input features of DNN.
Different from the traditional methods that resort to the optimization algorithm or the time-consuming simulations, DNN can straightforwardly learn to make the numerology selection from the channel characteristics without any approximations.
With the help of the proposed DNN, 
we can clearly display the decision boundaries and clarify the application scopes of different numerologies.

The remainder of this paper is organized as follows.
Section II briefly analyses the ISI and ICI caused by channel delay and Doppler spreads as well as provides the definition of SNR loss.
In Section III, we propose a DNN-based approach for numerology selection.
Numerical results and performance evaluation are presented in Section IV. Conclusions are drawn in Section V.

\section{System Model}
We consider an OFDM system with a candidate numerology set in which the values are assigned to 
the subcarrier spacing $ f_u $ and CP ratio $ \mu_u $.
The system bandwidth $ B $ is divided into $ N_u $ orthogonal subcarriers where an encoded
symbol is transmitted on each subcarrier.
The transmitted $ n $-th sample of the $ i $-th OFDM symbol using the $ u $-th numerology is
\begin{equation}\label{}
x_i(n)=\frac{1}{N_u} \sum_{k=0}^{N_u-1} X_i(k) e^{j 2 \pi k n / N_u}, n \in[-G_u, N_u-1],
\end{equation}
where $ X_i(k) $ is the complex-valued symbol at the $ k $-th subcarrier with unit power and $ G_u $ denotes the length of the CP.

The key idea of numerology selection is that the subcarrier spacing $ f_u $ and CP ratio $ \mu_u $ are adaptively determined according to the channel characteristics.
To facilitate the implementation, a commonly used numerology family is defined as 
$ f_{u}=2^{\kappa_{u}} f_{1} $,
where $ \kappa_{u} $ is an integer.
With this scaling approach, different numerologies can be implemented easily by different size of inverse discrete Fourier transform (IDFT) under the same sampling clock rate\cite{2017ICM-Multi}.
We consider the IDFT size $ N_u $ and the CP ratio $ \mu_u $ as the variable parameters in the candidate numerology set $ \mathcal{N}  $ which contains $ M $ combinations, i.e., $ \mathcal{N} :\{(N_1, \mu_1), (N_2, \mu_2),..., (N_M, \mu_M)\} $.
Then, the subcarrier spacing can be chosen by using different size of IDFT that leads to $ f_u = B/N_u $, and the CP length is changed by $ G_u=\mu_u N_u $ ( $ u=1,...,M $).

The channel in the time domain can be modeled as multipath Rayleigh fading channels where the time-varying channel impulse response $ h(n , l) $ represents the $ l $-th tapped delay line in the $ n $-th sample time.
The wide-sense stationary uncorrelated scattering (WSSUS) model \cite{P2003Mobile} is also used in this paper, 
in which different path delays are uncorrelated and
the channel correlation properties are invariant over time.
The delay spread of channel can be characterized by the PDP,
where the channel power of the $ l $-th tap is  $p_{l}=E\left[\left|h\left(n, l\right)\right|^{2}\right]$ and $ E\left[\cdot \right] $ denotes the expectation.
The maximum delay spread is $ L_{ch}T_s $, where $ T_s=1/B $ is the sampling time.
Then, we have $ 0\leq l\leq L_{ch}-1 $.
The total power is normalized as
$ \sum_{l=0}^{L_{ch}-1}p_{l}=1 $.

Meanwhile, because Doppler effect is usually different from path to path when signal arrives at the receiver, the maximum Doppler spread is used to describe the time-varying of channel, which is dependent on the mobile velocity and the carrier frequency. 
The maximum Doppler spread for a typical outdoor radio channel can be calculated straightforwardly as $ f_d = (v/c)F_c $, where $ v $ is the mobile velocity, $ c $ is the velocity of light, and $ F_c $ is the carrier frequency.

After the signal is transmitted through the channel,
the receiver drops the first $ G_u $ samples as CP and 
retains the following $ N_u $ samples to be processed.
The perfect synchronization of frequency and time is assumed and 
the received signal can be written as
\begin{equation}\label{}
y_{i}(n)=\sum_{l=0}^{L_{ch}-1} h(n , l) x_{i}\left((n-l)_{N_{u}}\right)+w_{i}(n),
\end{equation}
where $ w_{i}(n) $ is the complex-valued additive white Gaussian noise (AWGN) with zero mean and variance $ \sigma_0^2 $, and $ (\cdot)_{N_u} $ represents a cyclic shift in the base of $ N_u $.

After the DFT, the received signal in the frequency domain can be written as\cite{2004ITCE-ICI,2013ITWC-Optimal}

\begin{equation}\label{y}
\mathbf{y}_{{i}}=\mathbf{H}_{{i}}^{\mathrm{a v e}} \mathbf{X}_{{i}}+\mathbf{H}_{{i}}^{\mathrm{I C I}_{1}} \mathbf{X}_{{i}}+\mathbf{H}_{{i}}^{\mathrm{ISI}} \mathbf{X}_{{i}-{1}}+\mathbf{H}_{{i}}^{\mathrm{I C I}_{2}} \mathbf{X}_{{i}}+\mathbf{W}_{{i}},
\end{equation}
where $ \mathbf{W}_{{i}} $ is the AWGN in the frequency domain, 
the first term in the right-hand side is the desired signal without interference, the second term is the ICI caused by the time-varying channel, 
the third term is the ISI from the previous symbol and the fourth term is the ICI introduced by insufficient CP, respectively.
$ \mathbf{H}_{{i}}^{\mathrm{a v e}} $, $ \mathbf{H}_{{i}}^{\mathrm{I C I}_{1}} $, $ \mathbf{H}_{{i}}^{\mathrm{ISI}} $ and $ \mathbf{H}_{{i}}^{\mathrm{I C I}_{2}} $ are the corresponding $ N_u\times N_u $ matrices of channel effect with elements given by\cite{2004ITCE-ICI,2013ITWC-Optimal}


\begin{equation}\label{}
H_{i}^{\mathrm{a v e}}(k, k)=\frac{1}{N_{u}} \sum_{n=0}^{N_{u}-1} \sum_{l=0}^{L_{ch}-1} h(n ,l) e^{-\frac{j 2 \pi k l}{N_{u}}},
\end{equation}


\begin{equation}
\begin{split}\label{}
& H_{i}^{\mathrm{I C I}_{1}}(k, m)= \\
&\left\{ \!{\begin{array}{*{20}{c}}
	{\frac{1}{N_{u}} \sum_{n=0}^{N_{u}-1} \sum_{l=0}^{L_{ch}-1}h(n ,l) e^{-\frac{j 2 \pi k l}{N u}} e^{\frac{j 2 \pi n(m-k)}{N_{u}}},}\\
	{0,}
	\end{array}} \right.\begin{array}{*{20}{c}}
\!\!{k\neq m}\\
\!\!k= m
\end{array}
\end{split}
\end{equation}

\begin{equation}\label{}
\begin{aligned} H_{i}^{\mathrm{I S I}}(k, m)&= \\
 &\frac{1}{N_{u}} \sum_{l=G_{u}}^{L_{ch}-1} \sum_{n=0}^{l-G_{u}} h(n ,l) e^{-\frac{j 2 \pi k\left(l-G_{u}\right)}{N_{u}}}   e^{\frac{j 2 \pi n(m-k)}{N_{u}}} \end{aligned},
\end{equation}
and
\begin{equation}\label{}
\begin{aligned} H_{i}^{\mathrm{I C I}_{2}}(k, m)&=\\
&-\frac{1}{N_{u}} \sum_{l=G_{u}}^{L_{ch}-1} \sum_{m=0}^{l-G_{u}} h(n ,l)  e^{-\frac{j{2 \pi k l}}{N_{u}}} e^{\frac{j 2 \pi n(m-k)}{N_{u}}}, \end{aligned}
\end{equation}
where $ 0 \leq k \leq N_u-1, 0 \leq m \leq N_u-1 $.
Note that when the channel is time-invariant during a symbol duration, $ \mathbf{H}_{i}^{\mathrm{I C I}_{1}}$ is a full zero matrix.
Moreover, if the CP length is longer than the channel maximum delay, the delay spread will be perfectly eliminated, and we have $ \mathbf{H}_{{i}}^{\mathrm{ISI}}=\mathbf{0} \text { and } \mathbf{H}_{{i}}^{\mathrm{ICI}_{2}}=\mathbf{0} $.

From the analysis above, the desired power can be written as 
\begin{equation}\label{}
P_U = \frac{1}{N_{u}}\sum_{k=0}^{N_{u}-1}E\left[\left|{H}_{{i}}^{\mathrm{a v e}}(k, k)\right|^{2}\right].
\end{equation}
The total interference power is
\begin{equation}\label{}
\begin{aligned} 
P_I = \frac{1}{N_{u}}\sum_{k=0}^{N_{u}-1}E\left[\sum_{m=0, m \neq k}^{N_{u}-1}\left|{H}_{{i}}^{\mathrm{I S I}}(k, m)
+{H}_{{i}}^{\mathrm{I C I}_{2}}(k, m)\right|^{2}\right]\\
+\frac{1}{N_{u}}\sum_{k=0}^{N_{u}-1}E\left[\sum_{m=0}^{N_{u}-1}\left|{H}_{{i}}^{\mathrm{I C I}_{1}}(k, m)\right|^{2}\right].
\end{aligned}
\end{equation}

In order to evaluate the influence of ISI, ICI and noise power, we define the SINR at the output of the DFT as the ratio of the power of the desired component to the power of the interference, which can be denoted as 
\begin{equation}\label{SINR}
\mathrm{SINR}=\frac{ \frac{N_u}{N_u+G_u} P_{U}}{ \frac{N_u}{N_u+G_u}P_{I}+\sigma^2_{0}}=\frac{P_{U}}{P_{I}+({1+\mu_u})\sigma^2_{0}}.
\end{equation}
It should be noted that when the channel has no delay and Doppler spreads, the CP can be left out and the received SNR is represented by $ \mathrm{\gamma}=1/\sigma_0^2 $. 
Nonetheless, when considering the ISI and ICI, the resulting SNR loss is calculated as
\begin{equation}\label{SNRloss}
{\Delta\gamma}=10 \log_{10} \left(\frac{ P_{I}+ (1+\mu_u)\sigma^2_{0}}{P_U\sigma^2_{0}}\right).
\end{equation}
We can see from (\ref{SINR}) that if noise power $ \sigma_0^2 $ is small, the SINR will be limited by $P_{U} /P_{I}$, which demonstrates that the ISI and ICI impede the performance improvement as the increase of the transmitted power.
Furthermore, the expression (\ref{SNRloss}) shows that the noise power has significant impact on the selection of CP ratio.
When $ \sigma_0^2 $ is large, the SNR loss would be more sensitive to the value of $ \mu_u $.
Therefore, the waveform numerology should be carefully selected to reduce the interference according to the channel characteristics. 


\section{Proposed DNN-Based Numerology Selection Approach}

To minimize the SNR loss $ {\Delta\gamma} $, the numerology selection problem can be formulated as
\begin{equation}\label{opt}
\begin{array}{c}\min _{\left\{N_{u}, \mu_u\right\}}  \frac{ P_{I}+ (1+\mu_u)\sigma^2_{0}}{P_U\sigma^2_{0}} \\ \text { s.t. } (N_u,\mu_u) \in \mathcal{N}\end{array}.
\end{equation}
Due to the expectation operation in the objective, it is difficult to solve this problem.
Although the close-form expression can be approximately derived under specific channel assumption, it may still deviate the optimal solution.
In this work, we propose a fully connected DNN to address this problem in (\ref{opt}).

In order to learn effective representations from the channel,  
we construct a feature set that contains the channel noise power, the mobile velocity along with the channel PDP.
The PDP can be expressed as an $ L $-dimensional vector
\begin{equation}\label{delay}
\mathbf{p}_{h}=\left[p_{0}, p_{1}, \ldots, p_{L_{c h}}, 0, \ldots, 0\right]^T,
\end{equation}
where $ L_{ch}\leq L $.
The feature vector $ \mathbf{d}\in \mathbb{R}^{L+2} $ can be constructed by concatenating the noise power $ \sigma_0^2 $, mobile velocity $ v $ and the channel PDP vector $ \mathbf{p} $ as
\begin{equation}\label{}
\mathbf{d} = \left[ \sigma_0^2 , v, \mathbf{p}^T\right]^T.
\end{equation}
Then, 
the feature vector $ \mathbf{d} $ is used as the input of the network.

We use $ r $ to represent the number of layers where $ r=0,1,...,R $.
The number of neurons in each layer is $ D_r $.
Then, the $ (L+2) $-dimensional input layer has $ D_0=L+2 $ neurons and the output layer has $ D_R=M $ neurons that equal to the number of candidate numerologies.
The hidden layers utilize the rectified linear unit (relu) activation function.
The last layer applies softmax as the activation function to normalize the output of each neuron, which indicates the probability that the current channel characteristic is suited to the corresponding numerology.
We also apply the second norm regularization to alleviate the overfitting and the adam optimizer [?] along with the categorical cross entropy loss function for the network training.

The training samples can be generated by simulation or measurement.
When the observation time is enough for the reasonable expectation of interference,
the supervised learning is applied to train the networks by using the numerology with minimal SNR loss as the ground truth.
The training of the neural network is based on mini-batch stochastic gradient descent and backpropagation methods.
When the training is completed and achieves a good accuracy, we can deploy the proposed DNN for selecting the numerology.

\section{Numerical Results}

In this section, simulations are conducted to validate the performance of the proposed numerology selection approach.
In the simulations, we consider the communication system with 5 MHz bandwidth and 2 GHz carrier frequency.
The subcarrier spacing and CP ratio are chosen from the candidate set with 6 numerologies as:
\begin{equation}\label{}
\mathcal{N}:\left\{ {\begin{array}{*{20}{c}}
	{\begin{array}{*{20}{c}}
		{{N_1} = 240,{\mu _1} = 1/4};\\
		{{N_2} = 480,{\mu _2} = 1/4};\\
		{{N_3} = 960,{\mu _3} = 1/4};
		\end{array}}&{\begin{array}{*{20}{c}}
		{{N_4} = 240,{\mu _4} = 1/10};\\
		{{N_5} = 480,{\mu _5} = 1/10};\\
		{{N_6} = 960,{\mu _6} = 1/10.}
		\end{array}}
	\end{array}} \right\}
\end{equation}
The wireless channel is assumed to exhibit exponentially decaying PDP and the maximum delay spread is defined as the time at which the multipath power falls $ 20 $ dB below that of the strongest component.

To generate training samples, we use 2000 random channel conditions in which the rms-delay spread and mobile velocity are evenly distributed across  [0.2$ \mu s $, 10$ \mu s $] and [1 km/h, 500 km/h], respectively. 
Then, we simulate the transmission of OFDM waveform using all possible numerologies for different 5000 channel realizations, where the SNR changes from 0 dB to 49 dB with a 1 dB step.
In this way, the training samples are classified to the different labels of numerology 
and the class label is the index of the optimal numerology that can achieve the minimum SNR loss.

The proposed DNN model consists of five layers, three of which are hidden layers.
As the maximum delay spread of the considered channel conditions is no longer than the minimum symbol length of the numerologies, i.e., $ L_{ch}<N_1 $,
the PDP is extracted as an $ L=256 $ dimensional vector.
Concatenating the noise power, the mobile velocity and the PDP, we have a 258-dimensional input of DNN.
The numbers of neurons in hidden layers are 300, 200 and 100, respectively.
The output layer has 6 neurons corresponding to the indexes of the numerologies.

\begin{figure}[t]
	\centering
	\includegraphics[width=55mm]{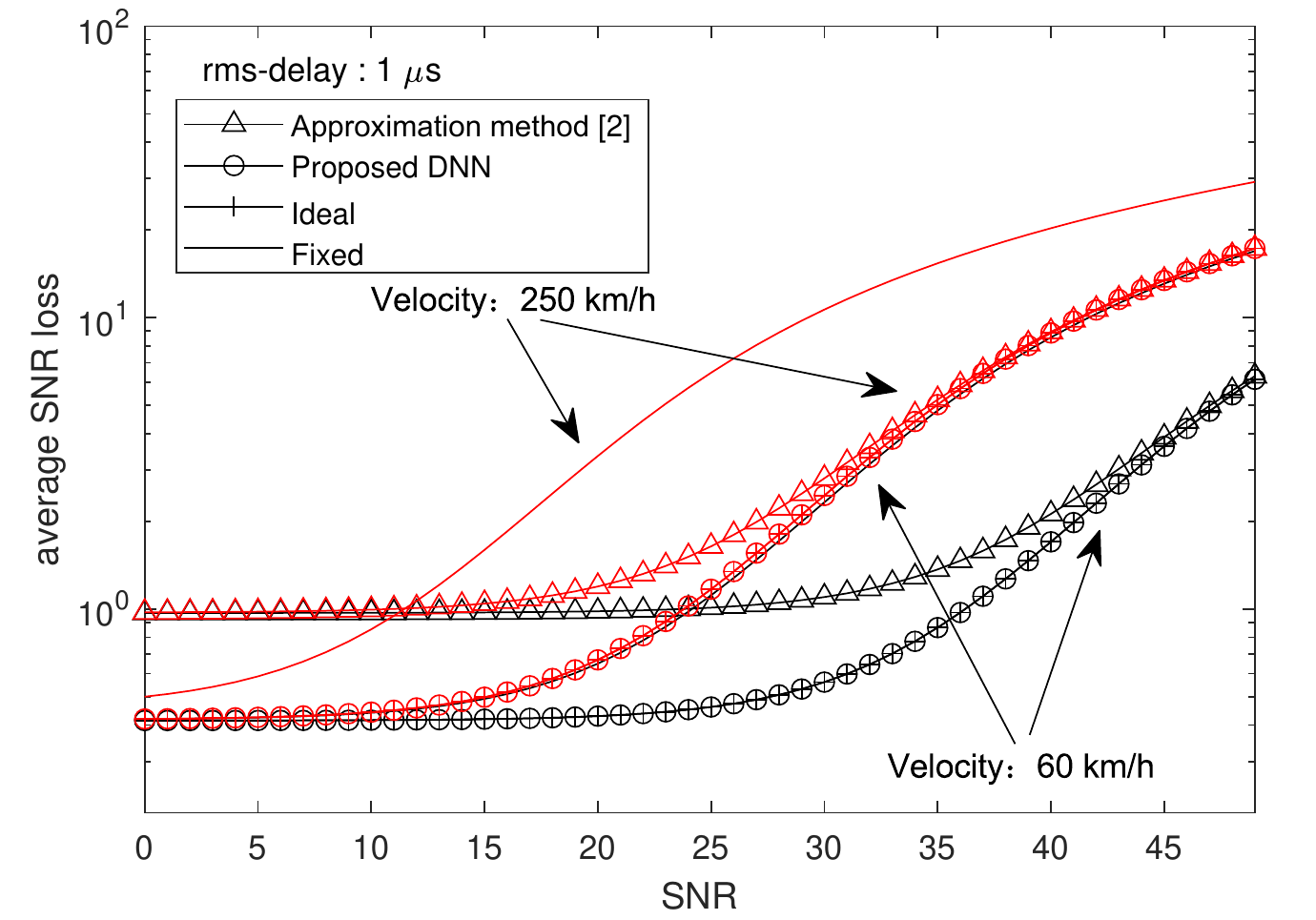}
	\caption{SNR loss of ideal results, proposed DNN and approximation method with channel delay 1 $ \mu s $ and mobile velocity $ 60, 250 $ km/h.}
	\label{figSNR1}
\end{figure}
\begin{figure}[t]
	\centering
	\includegraphics[width=55mm]{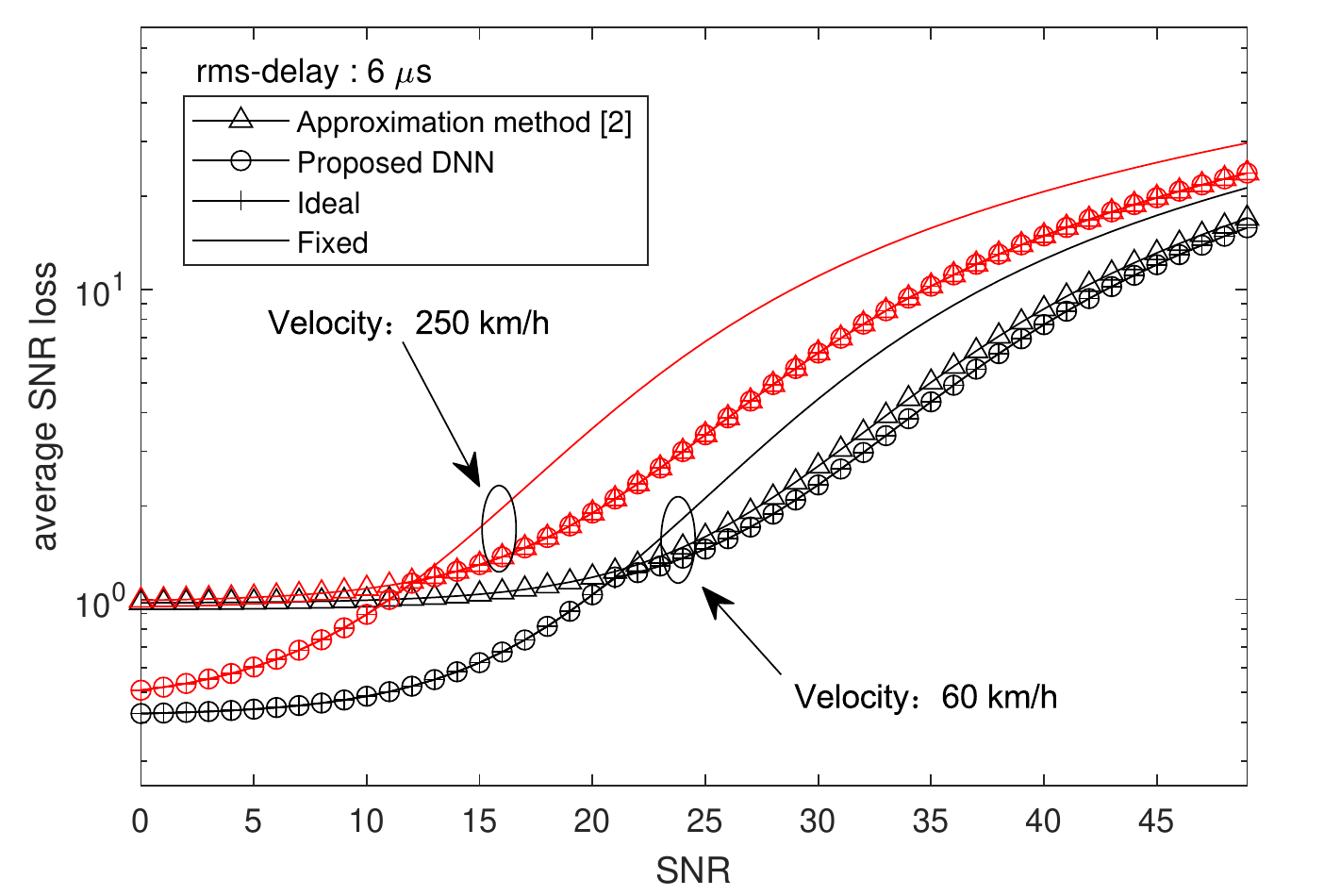}
	\caption{SNR loss of ideal results, proposed DNN and approximation method with channel delay 6 $ \mu s $ and mobile velocity $ 60, 250 $ km/h.}
	\label{figSNR2}
\end{figure}

In Fig. \ref{figSNR1} and Fig. \ref{figSNR2}, we depict the average achievable SINR of the ideal results, the proposed DNN-based numerology selection, the approximation method used in \cite{2013ITWC-Optimal} and the fixed numerology.
As can be seen, the SNR loss will increase when the mobile velocity is leveled up from 60 km/h to 250 km/h.
However, the proposed DNN can always achieve the same SNR loss as the ideal results under the corresponding channel conditions.
Furthermore, it can be found in Fig. \ref{figSNR1} that when SNR is high, the curves of SNR loss are very close between the proposed approach and the approximation method.
This is because the noise power and CP ratio are not considered by the approximation method and become insignificant in high SNR region.
Although the similar performance might be achieved in high SNR region, the performance degradation is still obvious in low SNR region.
This phenomenon is also shown in Fig. \ref{figSNR2} where the remarkable difference of SNR loss between the approximation method and proposed approach is displayed in high SNR region with $ v=60 $ km/h.
It can also be observed that when the SNR is larger than 12 dB and velocity is $ 250 $ km/h, the numerologies selected by the approximation method and proposed approach are the same and lead to the identical SNR loss.
In addition, the discontinuous point in the curve of the proposed approach is attributed to the change of numerology, which is associated with the variation of noise power.

\begin{figure}[t]
	\centering
	\subfigure[SNR = 5 dB]{\includegraphics[width=43mm]{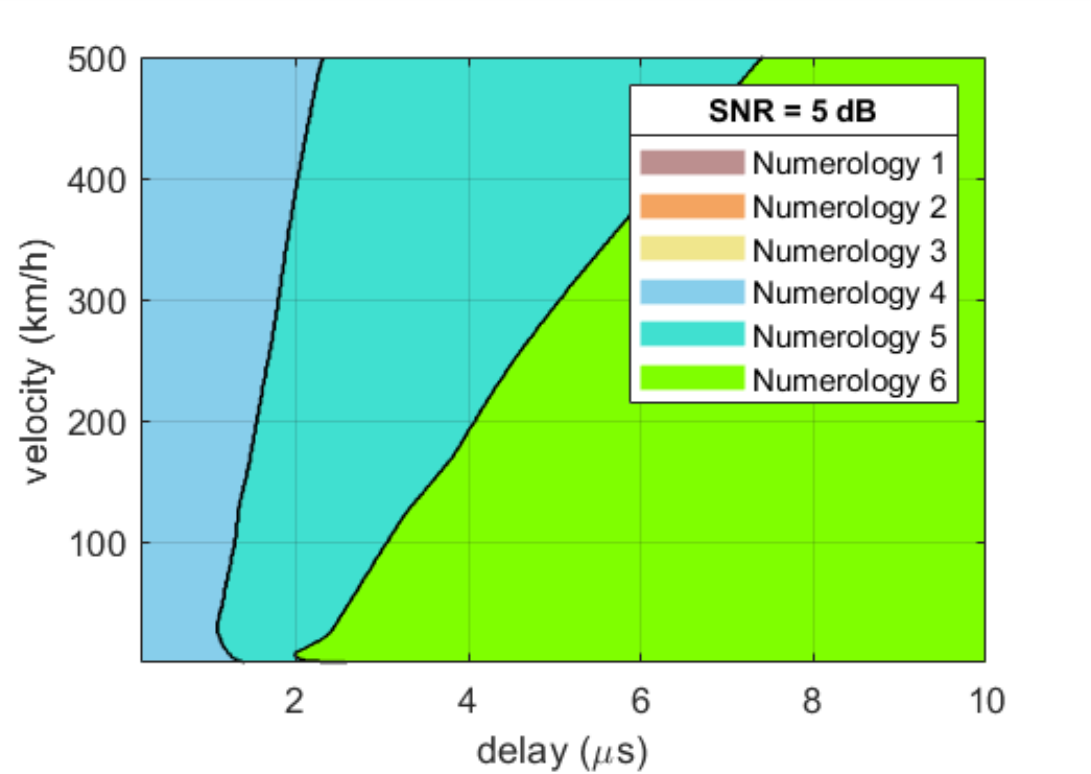}}
	\subfigure[SNR = 25 dB]{\includegraphics[width=43mm]{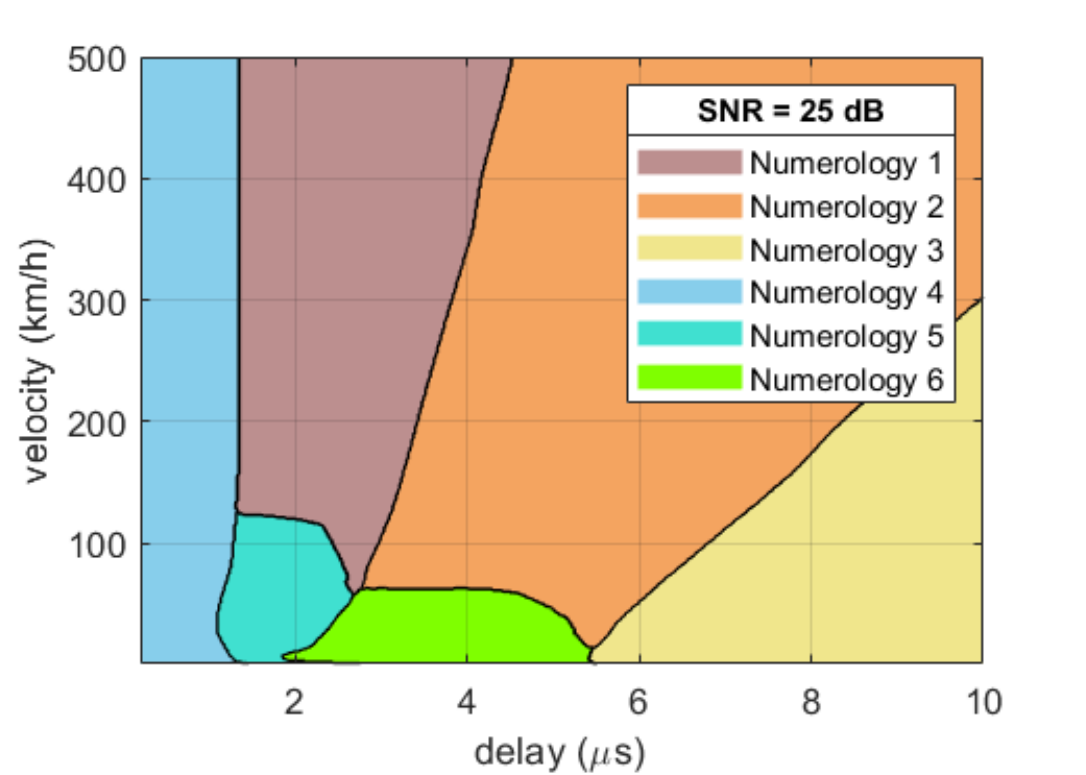}}
	\subfigure[SNR = 45 dB]{\includegraphics[width=43mm]{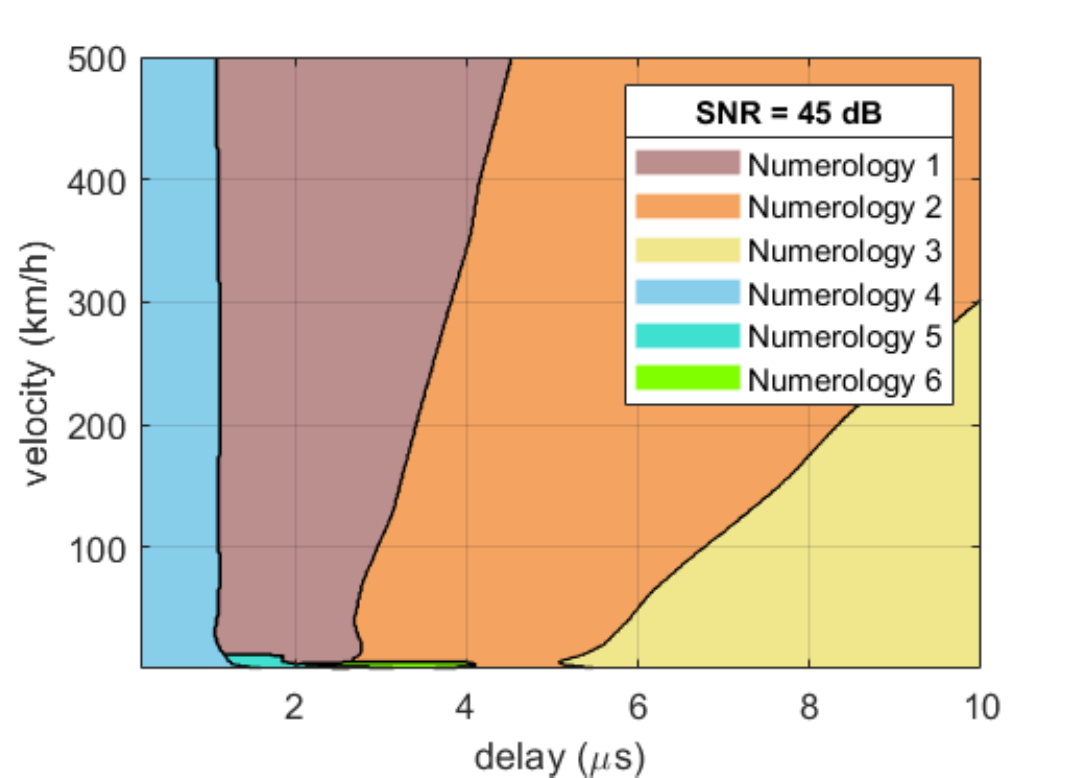}}
	\caption{ Decision boundaries of different numerologies under the considered channel conditions with SNR = 5, 25, 45 dB.}
	\label{fig3}
\end{figure}
%
%
The decision boundaries of numerology selection according to channel conditions at different SNRs are presented in Fig. \ref{fig3}.
The SNRs in three subfigures are $ 5, 25, 45  $ dB, respectively.
We can observe that when the SNR is 5 dB, only the numerologies with CP ratio of $ 1/10 $ are used.
On the other hand, as the SNR reaches at 45 dB, the numerologies with CP ratio of $ 1/4 $ are the major part of selected numerologies under the considered channel conditions.
It demonstrates that the SNR loss caused by CP overhead is dominant in low SNR region, while it will become negligible in high SNR region.
Furthermore, as the delay spread is enlarged, the numerologies with smaller subcarrier spacings are more likely to be selected.
The reason is that the smaller subcarrier spacing results in larger symbol duration and correspondingly longer CP to confront the delay spread.
Additionally, when the delay spread is relatively short, the ICI takes the dominant role of interference.
Thus, the large subcarrier spacing is preferred to withstand the time-varying channel.

\section{Conclusions}
In this letter, we present a DNN-based approach for numerology selection of OFDM systems.
The received signals suffer from the ISI, ICI and channel noise according to the channel characteristics.
Different numerologies are suited to different channel characteristics in terms of the interference level.
The proposed DNN is able to provide a reliable selection of numerology by using the PDP, mobile velocity and noise power as the input features.
Simulation results demonstrate that the proposed approach outperforms the existing methods in terms of the SNR loss and the application scopes of different numerologies for different channel characteristics are discussed.

\bibliographystyle{IEEEtran}
\bibliography{2020Spring}

\end{document}